\documentclass[11pt]{article}

\usepackage[vmargin=1.00in,hmargin=1.00in,centering,letterpaper]{geometry}

\usepackage{fullpage}
\usepackage[authoryear]{natbib}
\usepackage{hyperref}

\begin{document}

\title{\textbf{Interleaving Computational and Inferential Thinking: \\
Data Science for Undergraduates at Berkeley}}

\author{Ani Adhikari~\footnotemark[2] \quad \quad  John DeNero~\footnotemark[1] 
\quad \quad Michael I. Jordan~\footnotemark[1]~\footnotemark[2]\\
	\\
	\footnotemark[1]~~Department of Electrical Engineering and Computer Sciences\\
	\footnotemark[2]~~Department of Statistics\\
	University of California, Berkeley}

\date{}

\maketitle 

\begin{abstract}
The undergraduate data science curriculum at the University of California, Berkeley is
anchored in five new courses that emphasize computational thinking, inferential thinking, and the perspective gained by working on real-world problems.  We believe
that interleaving these elements within our core courses is essential to preparing
students to engage in data-driven inquiry at the scale that contemporary scientific and industrial applications demand.  This new curriculum is already reshaping the undergraduate experience at Berkeley, where these courses have become some of the
most popular on campus and have led to a surging interest in a new undergraduate major and minor program in data science.
\end{abstract}

\section{Introduction}

One of the most challenging---but ultimately most rewarding---ways to develop an undergraduate curriculum is to aim for the grand conceptual achievements of a field, stripping away the inessentials and conveying the core ideas in a way that reveals their beauty, their universality, and their contemporary relevance.  This has been done for many fields in the sciences and humanities, and the introductory courses in these fields have stabilized and stood the test of time.  Are we in an era in which this can be done for data science?
We have approached this question with cautious optimism at Berkeley.  The caution stems from a belief that the appropriate scope for data science education is a very broad one.  Indeed, we view data science as a form of liberal arts for the 21st century---a mingling of the computational and inferential disciplines that flowered during the 20th century and an approach to science and technology that permits empirical investigations at unprecedented scale and scope.  Given this vast remit, the challenge of perceiving and distilling a curricular core seems daunting at best.
And yet we are optimistic.  We view data science as a phenomenon that took shape slowly and steadily over the past century, with the flowering of its various parts arising from a common root.  Indeed, during the gestation of the computational and inferential disciplines at the beginning of the 20th century, individuals such as John von Neumann, Andrey Kolmogorov, Alan Turing, Jerzy Neyman, Norbert Wiener, and Abraham Wald blended the deductive and inductive traditions of mathematics to devise rigorous formulations of concepts such as ``algorithm,'' ``probability,'' 
``inference,'' ``feedback,'' ``uncertainty,'' ``model,'' and ``risk.''  
These formulations provided new ways to think about data and its role in scientific inquiry.  The scope of these developments was further broadened in the hands of individuals such as David Blackwell, Claude Shannon, and Herbert Robbins, whose work led to new perspectives on economics, communications, and psychology.

As the 20th century progressed the unity began to be less apparent.  From the original foundations, major new academic disciplines emerged---computer science, control theory, information theory, signal processing, and mathematical statistics---each focused on particular aspects of the overall set of challenges associated with information, inference, and decisions.  Whereas von Neumann, Kolmogorov, et al.\ would likely have resisted being labeled with a single one of these disciplinary labels, subsequent researchers have generally pursued their careers entirely within a single discipline.

Data science has brought the original threads back together.  Data science focuses on real-world problems involving data and decisions.  That these problems are not generally the province of a single discipline is something that students are prepared to accept without much debate.  They can appreciate the need to go beyond the mere processing of data and call forth the broader set of ideas: the specification of inferential goals, the development of models that aim to capture the way in which data may have arisen, the crafting of algorithms that are responsive to the models and goals, an understanding of the feedback mechanisms that affect the data and the interpretation of the results, concern about uncertainty and risk, and concern about the human implications of automated data analysis and decision-making.  At Berkeley we have coined a phrase---``computational thinking and inferential thinking''---to capture one important aspect of our vision for a data science curriculum.  ``Computational thinking'' is the goal of a modern introductory class in computer science, where the focus is on notions of abstraction, modularity, and efficiency.  ``Inferential thinking'' is the goal of a modern introductory class in statistics, with a focus on ideas such as populations, sampling, Bayes theorem, causality, and robustness.  Placing the two in juxtaposition brings together many of the conceptual achievements of the past century referred to earlier.

Moreover, such a juxtaposition profits from the complementary problem-solving styles associated with computer science and statistics.  Computer science has a ``builder'' spirit associated with it.  Students who write computer programs feel empowered. They are not merely learning a formalism, but they are creating working artifacts.  Statistics, on the other hand, embodies a ``collaborator'' spirit.  Statisticians learn to embed themselves in teams along with domain experts and contribute to the conceptual flow of a project. Putting these two styles together is a natural, desirable feature of a data science curriculum that aims to blend computation and inference.

In the design of our curriculum, we have also included ``real-world implications'' as a third foundational leg, to highlight the fact that while data science may have its roots in mathematical formalism, it is crucially a real-world enterprise.  Here too there are important historical precursors that have inspired us.  Individuals such as John Tukey and Leo Breiman, trained as mathematicians, came to emphasize the open-ended, exploratory nature of data analysis, and the necessity of trying things out on real-world data.  This perspective complemented the formal mathematical perspective that dominated academic research communities, bringing difficult-to-formalize, but essential, notions such as visualization, interpretability, and criticism to the fore.  Another key historical reference is the database researcher Jim Gray, whose foundational work on systems for indexing and querying massive, evolving collections of data led him to envisage an emerging ``fourth paradigm'' of highly collaborative, data-intensive science.  Finally, another powerful historical thread that has influenced us comes from social science, where the \emph{contextual} nature of data and empirical investigation is a major theme.  We've been influenced by the work of social scientists Sheila Jasanoff and Donna Haraway, and a particular inspiration has been the writing of Ursula Franklin, whose wide-ranging commentary on technology as a complex system is exemplified by the following quote: ``Technologies are developed and used within a particular social, economic, and political context. They arise out of a social structure, they are grafted on to it, and they may reinforce it or destroy it, often in ways that are neither foreseen nor foreseeable''~\citep{Franklin}.  We have accordingly aimed to strike a tone of respect for consequences, foreseen and unforeseen, in our teaching of data science.

There was indeed an interesting unforeseen consequence that arose when we first began to teach the new courses.  We found that whatever the notion of ``real world'' might mean for us, students had their own ideas.  Students come to our curriculum with their own questions and passions, providing a vivid example of the notion of ``context.''  We came to realize that ``real world'' is best viewed as a student-centric concept---it can be whatever a given student wishes for it to be.  A data science education can aim at empowering students to not merely solve others' problems, but to solve their own problems.   A data science curriculum can empower them to find data that are relevant to their questions and to provide convincing analyses of those data.  The word ``convincing'' is important here---a good data scientist should be able to convince not only himself or herself of an analysis, but should be able to convince others as well.

In short, our learning goals for the new curriculum were multi-fold, and avidly cross-disciplinary: we wanted students to understand how to formulate meaningful inferential problems, collect data relevant to those problems, build data analysis pipelines that allow problems to be solved at a range of scales, carry out analyses that are convincing, and make assertions or policy recommendations that carry weight.  Moreover, throughout this process we wanted students to be attentive to the social, cultural, and ethical contexts of the problems that they are formulating and aiming to solve, and we wanted to empower students to pursue their own unique perspective as data scientists.  These goals echo many of those proposed in recent efforts to reform undergraduate statistics curriculas~\citep{cobb2015}.  They also have some elements in common with proposals that build new data science curricula by drawing carefully from existing curricula in computer science, statistics, and mathematics, while providing experience with real-world data~\citep{deveaux} but our emphasis is different, and where we particularly align with the latter authors is when they state, ``We believe that many of the courses traditionally found in computer science, statistics, and mathematics offerings should be redesigned for the data science major in the interests of efficiency and the potential synergy that integrated courses would offer.''  A thoroughgoing redesign is precisely what we have engaged in at Berkeley.

To turn high-level desiderata into an actual class sequence, we found it helpful to start from scratch.  We first focused on first-year students, and we avoided making any strong assumptions about mathematical background or programming skills.  Our thought was that if ``computational thinking and inferential thinking'' is a powerful new force on the academic landscape, then its power should be evident without a great deal of background or formalism. Later courses then reinforce and expand the material by revisiting many of the same problems to which the students were exposed in their first course using more developed mathematical and computational tools and contextual frameworks.

\subsection*{Vignettes}

Let us give a concrete example of what it means to teach ``computational thinking and inferential thinking'' in an interlaced fashion.  A variant of this particular example can be found in our freshman-level course (Data 8, which we discuss in more detail below).

In general, concrete examples helped to shape our design of data science syllabii.  Rather than simply trying to assemble a syllabus as a mash-up of computer science topics and statistics topics, we found it more helpful to focus on specific problems and bring computing and statistics to bear jointly in the solution of the problem.

Consider the problem of deciding whether the juries in a given city are representative of the population of the city.  In particular, suppose that we have two columns of numbers, one listing the values of a certain demographic measurement for a set of jurors and the other one listing the values of the same measurement for a census of the population.  For concreteness, suppose that we have two columns of numbers with 100 elements in each column.  These columns will surely differ in various ways, but are these differences systematic, and reflective of some form of bias, or would the difference be expected by chance alone?

Rather than heading immediately down the traditional path of means, standard deviations, and t-tests, we treat this problem as an opportunity to focus on ``randomness.'' Recalling that we are at the freshman level and that we are not assuming knowledge of things like probability distributions or independence, we introduce the histogram as a stand-in for the notion of a distribution.  This requires an assumption, one that we make explicit: We assume that the data in each column are \emph{exchangeable}---meaning that the ordering of the points in the column is arbitrary.  From this assumption it is reasonable to reduce a column of numbers to a histogram, a visualizable data structure that records the proportion of the overall data set that falls in each bin.  Using the same bins for the two columns, we next introduce the idea of comparing two histograms. To do this, we simply take the difference between the two values in each bin, compute the absolute value, and sum up these absolute values across the bins.  A graduate-level researcher will recognize this sum as the \emph{variation distance}, and may be surprised that we are introducing such an advanced concept in a freshman course, in lieu of means and standard deviations.  But our students don't seem to find this concept to be difficult or unnatural.

For a particular set of data, suppose that the variation distance is 0.7.  Could this number have arisen by chance alone?

We're at a key juncture in inferential thinking.  The language of ``could have arisen'' suggests that we're asking a question that is not to be found by mere inspection of the data.  Some thinking is needed.  We introduce the idea of a \emph{hypothetical}---a ``null hypothesis''---that there is actually no difference between the jurors and the population.  Entertaining this null hypothesis as a thought experiment, we can simply lump all of the data together into a single column of 200 numbers.  Recalling our exchangeability assumption, we are also free to permute this column---the order doesn't matter.  For each permutation, we then split the column into two new columns of 100 numbers
each---for example, we take the first 100 numbers and the second 100 numbers.  We then form two histograms from these columns and compute the variation distance.  What we are doing here is simulating the randomness that we would expect if juries are selected in a way that accords with our thought experiment.  We've operationalized the notion of ``by chance alone.''

This process yields a collection of numbers, from which we form a new histogram.  We give this histogram a name---it is the \emph{sampling distribution}.  It captures differences between histograms due to chance alone.  We now reconsider that value of 0.7.  If it is in the center of the sampling distribution, then it would have been expected under the null hypothesis.  We have no reason to reject the null hypothesis.  If, on the other hand, it is in the tail of the histogram, then it is ``surprising.''  Although we could decide that something surprising has happened, it is perhaps more plausible to consider that we were in fact wrong to tentatively assume that there was no difference between the juries and the population.  We reject the null hypothesis.

This line of argument has introduced the idea of a \emph{permutation test}, an idea that goes back several decades~\citep{Fisher, Pitman}.  The test embodies core inferential notions of exchangeability, a counterfactual null hypothesis, random sampling, and the logic of ``surprise.''  The concreteness has the virtue that these core notions are laid bare, not cluttered by ideas such as standard deviations, Gaussian distributions, or (horrors) asymptotics.

The concreteness also has the virtue that an inquiring freshman mind can immediately see various weaknesses and hidden assumptions in our setup.  Caveats are needed in interpreting the results of such a procedure.  Thus, with little prompting, students are able to take a critical perspective on the ideas that they have learned, particularly when those ideas are confronted with the messiness of the real world.  Moreover, they can take their new machinery out into the real world---working with data sets and scientific questions that they find interesting---and to consider those caveats in a setting where there are consequences.

Where is the computational thinking in this?  One possible answer is that it is present throughout, in that we ask students to work with real computer code (Python in our case) as they form histograms, compute variation distances, and form sampling distributions. Understanding the mechanics of the program in which the inferential process is expressed provides a solid foundation for conceptual understanding.  There is also an opportunity to relay a deeper lesson about abstraction.  We ask students whether a similar analysis could be carried out if the census were not available, and instead they had only the relative frequencies in the city population for various ranges of demographic values. Instead of comparing two columns of numbers describing individuals, they now must compare a column describing individual jurors with a table describing population frequencies.  Students can affirm for themselves that the same reasoning applies by identifying that histograms are being compared in their hypothesis test, rather than columns of numbers, and that a histogram can be produced from either a column of numbers or a table of bin frequencies. A small change to the simulation, motivated by the idea that the meaning of a histogram is independent of how it was computed, allows this inferential technique to apply to a new data condition. Similar reasoning generalizes the technique to categorical variables, where bar graphs take the place of histograms and sound judgment can be made about whether a difference exists between two groups without ever mentioning chi-squared distributions.

There is also an opportunity to teach deeper computational concepts here.  How, from a computational point of view does one (randomly) permute a list of numbers?  One possibility is to randomly swap each pair of numbers.  But we note a key problem---that the \emph{computational complexity} of this algorithm is quadratic in the length of the list.  Such an algorithm will not perform well on large-scale problems.  Is there a better alternative?  The answer is ``yes,'' and it is given by \emph{Knuth's algorithm}---walk through the list in order and randomly swap the $k$th element with one of preceding $k-1$ items.  
Of course, the goal at this point is to have one of the students proposing this algorithm, to have another student noticing that it has linear complexity, and to have all of the students learning the basic concepts in Python that are needed to implement the algorithm.

Students engage with enthusiasm when real-world problems are expressed with real data and described in the context of real decisions.  For the jury problem, we make use of data collected by a 2010 ACLU study comparing the demographic composition of jury panels to the county from which they were drawn~\citep{aclu2010}.  At the same time that students are combining computational thinking and inferential thinking to make convincing statements about whether a jury is representative of the population from which it was drawn, they are exposed to an example of the role of data analysis in the formation of legal precedent and government policy.

We refer to such an interlacing of computation, inference, and a real-world problem setting as a ``vignette.'' In our design of a data science curriculum it was helpful to generate many such vignettes.

We now turn to an overview of the five courses\footnote{Course materials are publicly
available at \href{http://data8.org}{data8.org}, \href{http://ds100.org}{ds100.org}, \href{http://prob140.org}{prob140.org}, \href{https://data102.org}{data102.org}, 
and \href{https://data.berkeley.edu/data-104-human-contexts-and-ethics-data}{data.berkeley.edu/data-104-human-contexts-and-ethics-data}.}
that form the backbone of the Berkeley data science curriculum.  They are: (1) Data 8, a flagship freshman-level course that introduces Python programming, data manipulation, and basic statistical inference; (2) Data 100, a sophomore/junior-level course that introduces data-analysis pipelines and gradient-based algorithms for fitting statistical models; (3) Data 140, a sophomore/junior-level course that begins to assemble the formal tools of probability theory, doing so from a computational perspective; (4) Data 102, a senior-level course that focuses on the decision-making side of inference, including false-discovery rate control, bandit algorithms, optimal control theory, differential privacy, and matching markets; and (5) Data 104, a junior/senior-level course that provides conceptual frameworks for thinking through the implications and ethics of data collection, data analysis, and algorithmic decisions. Throughout, an integrated set of labs, homework assignments, and programming projects help to keep the focus on real-world problems.  The focus on real-world problems also makes it natural to connect to a wide range of other disciplines.

\section{Data 8: The Foundations of Data Science}

Starting from a blank slate, or more aptly a blank screen, we endeavored to build a course about how data-driven reasoning is properly carried out now that programmable computers are the central tool of statistical inquiry.  We sought to design a course that would integrate all that had been previously discovered about how best to introduce students to both statistics and computer science, but also distill the essential lessons learned by the many fruitful cross-disciplinary collaborations between researchers in statistics, computer science, and neighboring disciplines that gave rise to the emerging discipline of data science.  Indeed, our experience in graduate-level research and teaching was an essential inspiration for the design of Data 8, our freshman-year course in data science.

The focus of the course is on reasoning, visualization, and interpretation, rather 
than calculations or the use of software packages.  This approach is inspired by 
the boldly innovative \emph{Statistics} by \citet{fpp}, a textbook that transformed the way the field of statistics was introduced to 
undergraduates at Berkeley and around the world.  We emphasize the importance 
of consistency between the assumptions in a statistical model and the way in 
which the data are generated; we discuss what can and cannot be justifiably 
concluded based on our calculations; and we try to express mathematical statements 
as precisely as possible in plain English. 

\subsection{Computation}

However, the presentation of these classical lessons changed sharply when we adopted computation as a central tool. Unlike existing introductory statistics courses, examples in Data 8 do not begin with data summaries in the form of one or two graphs or a few numerical values. Rather, we give students the entire data set and teach them to generate the graphs or summary statistics that will become the centerpieces of their statistical arguments. When students write the code to generate a visualization, they pay attention to the choices of scales and understand the individuals on whom the data were measured. They can experiment with different choices. They are thus empowered to ask and answer their own questions in addition to the ones discussed in class.

Instead of restricting themselves to arithmetic that can be easily done on hand calculators, our students build full Python programs as Jupyter Notebooks that first format the data, then summarize, visualize, and quantify differences using randomized simulation as the problem requires. The freedom to explore a problem using a computer is transformational.  Few students have ever been devotees of hand calculators. For many of them that calculator emerged from their backpacks only for statistics exercises and for nothing else.  By contrast, a web browser is an indispensable part of the 21st-century-student's daily existence.  When students open up a browser-based Jupyter notebook in a data science class, that action is as natural to them as breathing, regardless of their background or academic interests.  Data 8 takes full advantage.

The ability to program requires both conceptual understanding and practice, yet Data 8 does not have a programming prerequisite, and close to half of the students who enroll have never programmed before.  This design choice was intentional.  The motivation of typical introductory computer science courses is to build software applications, rather than to use computation as a tool for inquiry.  A data science course without a programming prerequisite offers students an alternative motivation for learning to program, as a means to understand the world.  This motivation aligns to a broad set of student interests that are often complementary to those targeted by existing computer science courses. Students recognize this---typical evaluations of Data 8 include ``helpful for all majors'' and ``the best opportunity to get your hands on some coding techniques without any coding background.''
The focus of the programming content in the course is on composition and abstraction, rather than the many features of modern programming languages and data analysis packages that could easily consume months of instructional time.  This approach is inspired by the book \emph{Structure and Interpretation of Computer Programs}, by \citet{sicp}, 
another transformative classical textbook.  We highlight how a small set of primitive operations can be composed to carry out a wide range of data manipulation and visualization tasks; how simple built-in data structures can be used to represent a wide range of data scenarios; and how names and functions can be used to create compact and extensible programs that people can follow and understand.

Data 8 provides a great deal of practice using an essential but small and highly accessible subset of the Python programming language, with the goal of giving students sufficient experience with basic programming topics to achieve mastery.  Programming can be engaging even when limited to its most basic components---variables and procedure calls---when used to understand real-world data sets. For example, just four weeks into the course, Data 8 students carry out an extended analysis of world population growth over time, along with trends in fertility and child mortality rates, to form an argument based on data regarding whether the Earth's population will eventually stabilize or grow without bound.\footnote{The data and many of the visualizations for this assignment are from \href{https://gapminder.org}{GapMinder}.} This project does not require any iteration or control; just assigning variables, defining functions, and invoking methods.

To achieve this simplicity, Data 8 uses a Python module built for the course 
called \texttt{datascience}\footnote{See \href{http://data8.org/datascience}{data8.org/datascience} for 
documentation.} designed to ensure that students can carry out a wide range of 
table manipulation and data visualization operations using only a core set of 
the most fundamental programming language concepts.  Later parts of the course 
introduce control statements, but programming topics such as defining 
classes that are standard in introductory computer science courses are never covered in Data 8. Therefore, when students subsequently take an introductory programming course, the content is not redundant with Data 8, but the students who came through Data 8 are primed to recognize how the techniques they learn in later computing classes might apply to data problems.

\subsection{Inference and Prediction}

The arc of Data 8 motivates each programming topic with a new inferential goal.  In the 
first section of the course, students learn to program while learning about data 
visualization and summarization. Many questions can be answered with a carefully 
crafted picture. However, when two quantities differ and random sampling is involved, 
students naturally ask whether that difference is a real property of the world or 
just a quirk of the data.  Making this decision motivates the second part of the 
course.

To make such a decision, it is necessary to understand sampling variability. 
In traditional classes students have to imagine how the sample might have come 
out differently, and then either do some mathematics to quantify sampling variability 
or (as is common in introductory classes) accept and memorize some variance formulas. While demonstration by simulation is used in many classes, the students' work most commonly involves the use of standard deviations that they don't understand well. In Data 8, the students' primary tool for inference is to draw new samples and see the difference. Sampling variability is therefore expected and visible instead of being mysterious or obscure. 

Inference in Data 8 centers on simulation and the fundamental observation that 
the empirical distribution of a large random sample typically resembles the
distribution of the population from which the sample is drawn. Simulation takes 
three main forms:
\begin{itemize}
\item Simulating a multinomial random variable when the probabilities of all the classes are completely specified;
\item Random permutations of a pooled sample, for inference about the underlying distributions of the individual samples;
\item The bootstrap, to create new samples from a single large sample.
\end{itemize}

The third and final part of the course on prediction guides students to understand how machine learning systems are trained, applied, and evaluated without the conceptual overhead of optimization or calculus. Using simple linear regression and k-nearest-neighbor classification, students make predictions using real-world data sets. Although the prediction techniques are simple, the students' statements about the problem they are solving can be quite sophisticated at this stage, for example including confidence intervals around the estimated accuracy of their classifiers.

Using real-world data sets throughout naturally promotes discussion of the context in which data were collected and the social implications of collecting and analyzing data. Students explore the privacy implications of license plate scanners, the science behind estimating the age of the universe, and the complicated evidence linking dietary cholesterol to coronary heart disease.

\subsection{Modifications and Student Response}

Though Data 8 is largely the same now as in its pilot offering in Fall 2015, there have been some changes based on faculty and student response. Students in the pilot were exceptionally confident and curious, willing to take a chance on a new course based only on an interesting course description. With them, we were able to go further than we now can with a much more diverse group. For example, multiple regression has been removed from the syllabus.

The treatment of probability theory too has been cut down, focusing on shapes of distributions and uniform random sampling from finite populations. With a minimal use of notation, the course covers partitioning and addition, intersections and conditioning, and Bayes' theorem including a discussion about base rates. Probability calculations based on finite outcome spaces provide interesting settings for using array operations and iteration, but in the interests of time there are fewer of these now than there used to be. There is also less time spent on symmetry in random permutations. There is no formal treatment of random variables and distribution families other than the normal. However, Data 8 gives students an understanding of standard deviation and its relation to the normal curve, the variability of a random sample mean, and the Central Limit Theorem. This approach closely follows that of \citet{fpp}.

Students appreciate Data 8. Each semester's class consists of students of all years and dozens of different majors, the vast majority finding the course well worthwhile. ``This was easily the most influential and enjoyable class I've taken at Cal,'' said one. ``This was an extremely applicable course,'' said another, ``Because it opened up students' eyes on the widespread use of data science and how we can see it and use it in everyday life.'' Others made connections to their future work, as in, ``Companies look for people who are technologically aware and capable, therefore I believe it is an extremely beneficial class,'' or ``I feel I learned a lot and gained many skills I will take into my future work and academic career.'' See also Uptake and Engagement below.

\subsection{Connector Courses}

Accompanying Data 8 is a suite of ``connector'' courses designed to give students a 
more immersive experience in the data science of a particular domain.  These courses 
address the diverse interests and goals of the students in Data 8, provide students 
with engaging introductions to many different fields, and offer students a collaborative 
research-style environment in which to apply what they have learned in Data 8.

Connector courses require anywhere between 50\% to 100\% of the hours spent by students in a typical course, and the class sizes are typically small by Berkeley standards. Some connector courses have additional prerequisites. Data 8 is a prerequisite or corequisite for all connector courses. Students may take multiple connector courses, while they are taking Data 8 or in any future semester. 

Two of the connector courses are intended for students who want to go deeper into 
the computational or statistical underpinnings of Data 8. The Computer Science connector, 
\emph{Computational Structures in Data Science}, develops fundamental concepts of 
computation in the context of data analytics and, along with the programming in 
Data 8, provides students with the equivalent of a one-semester introductory CS 
course. The Statistics connector, \emph{Probability and Mathematical Statistics 
in Data Science}, provides students with a theoretical foundation that complements 
Data 8 and prepares students for higher-level classes in Statistics, Economics, 
and other fields. Students are eager for this preparation; each of these connectors is taken by hundreds of students every semester.

But the typical connector course is a more intimate project-based introduction to data science 
in a domain of application.  Class structures vary across connectors and fields. 
One common structure consists of a weekly two-hour session that is a combination 
of lecture and in-class lab time.  Students are required to complete a weekly lab, 
along with a more long-term project or assignment.

Departments all across campus have become enthusiastic partners in the connector 
project.  Examples of connector courses and their host departments include, among many others:

\begin{itemize}
\item \emph{Data Science and the Mind} [Cognitive Science]
\item \emph{How Does History Count? Exploring Japanese-American Internment through Digital Sources} [History] 
\item \emph{Data and Ethics} [Information] 
\item \emph{Immunotherapy of Cancer: Success and Failure} [Molecular and Cell Biology] 
\item \emph{Reading and Writing in the Digital Age} [English] 
\item \emph{Children in the Developing World} [Public Health] 
\item \emph{Data Science Applications in Physics} [Physics] 
\item \emph{Data Science for Smart Cities} [Civil Engineering] 
\item \emph{Data Science and Immigration} [Demography] 
\item \emph{Exploring Geospatial Data} [Environmental Science, Policy, and Management] 
\item \emph{Crime and Punishment: Taking the Measure of the US Justice System} [Legal Studies]
\item \emph{Data Science for Social Impact} [Sociology]
\end{itemize}

Through the connector courses, which have been offered ever since Data 8 was introduced in Fall 2015, data science education has become a campus-wide effort and is not restricted to a small number of STEM departments.  Connector courses embody Berkeley's view of data science as a way of thinking about the world, and are a formative element in our students' perception of data science as accessible
and interesting to all.

For faculty, connector courses offer many benefits. They are an unusual and effective way of attracting students to a field.  They provide a low-stress opportunity for faculty to expand the role of data science in their teaching and research, and sometimes also in their departments' degree programs. They create a community of faculty who are engaged in incorporating data science into their fields, in varying degrees and in different ways to be sure, but all working towards a better understanding of the value of data science in their domain.

Data Science Undergraduate Studies (DSUS) at Berkeley provides considerable support for faculty in connector courses and has produced a publicly available Guide for Instructors that aids in the development of such courses.  DSUS staff help faculty with computational resources and infrastructure, identifying qualified student assistants, and other logistical and developmental aspects. Support is also extended to faculty who wish to include smaller data science ``modules'' in other classroom settings, expanding the community of engaged faculty and the range of student experiences of data science. There are two annual summer faculty workshops, one for Berkeley faculty and one for faculty from other schools, on developing courses that draw upon the content and pedagogy of Data 8.  There is also a day-long workshop for graduate students on data science pedagogy.  During the semester, weekly meetings of connector instructors and support staff provide an opportunity for connector faculty to share their experiences and best practices as well as get the support they need for the smooth day-to-day running of their courses.

\section{Data 100: Principles and Techniques of Data Science}

While Data 8 students master the manipulation of small, tidy data tables that describe uniform random samples, Data 100 directly confronts the reality of modern data science: unstructured text, high dimensional observations, stratified samples, data sets that exceed the size of a program's working memory, and other twists that challenge students to generalize what they have learned to new settings. But at this stage they are far more capable due to the mathematics, statistics, and computer science courses they have taken since they started in Data 8. Data 100 draws connections across programming, linear algebra, calculus, and probability in order to expand the set of visualization, inference, and prediction problems that students can address. The course is designed to address the full lifecycle of a data science project using real data and real tools. One student summarized the course as, ``The class provides real-world data analysis skills that are highly prized by industry and prepares students to handle data at large scale.''

Like Data 8, Data 100 is organized around vignettes that expose the need for coordinating computational and inferential ideas in the context of real-world problems.  Inspired by one course project about analyzing Twitter data, a group of students in the course went on to build a service dedicated to identifying Russian Twitter bots. In another project, the students develop a system to predict the duration of a taxi ride in Manhattan based on historical records.  Data representation and interpretation are central concerns. Students begin with exploratory data analysis. Why were there so few taxi rides on January 23, 2016?  A record 27 inches of snow blanketed the city. Next, they must determine which taxi rides originated in Manhattan given geographic coordinates and a set of polygons that outline the island, and their implementation must efficiently process a large dataset. Only then can they turn to prediction and reflect on the results. Nothing about such a problem is a straightforward application of mathematics, computer science, or statistics.  As with many real-world problems, success requires a combination of applying general theory about making predictions with particular attention to details of the data and context. 

Data 100 exposes students to a variety of new computational topics. The course transitions students from the \texttt{datascience} module to 
\href{https://pandas.pydata.org/}{\texttt{pandas}} 
so that they can take advantage of Python's vast data science ecosystem. The programming concepts necessary to master \texttt{pandas}---slicing, class hierarchies, property methods, Boolean arrays, multiple views of an object, mutation, and type dispatching---are all covered in prerequisite programming courses (although not in Data 8). Therefore, introducing pandas is quick and efficient in Data 100, and students can appreciate the library's design and its interaction with Python's visualization and prediction libraries. But Data 100 also strives to make clear that Python is neither a necessary nor sufficient platform for data science. Interacting with a database management system is a substantial topic in the course, extending the basic SQL coverage in prerequisite courses by addressing import and export, data cleaning, types of joins, and sampling. The course projects demonstrate that leveraging the full capabilities of a database can greatly increase the feasible scale, efficiency, repeatability, and clarity of a data workflow. The intent of this approach is to ensure that students are prepared to apply what they have learned in external settings. One wrote, ``[This course] is useful for my research, and I could see myself using most of the topics in real life.'' Another wrote, ``I learned a lot about how data science might work in industry and going beyond Data 8 level of inference/intuition for analyzing data sets.''

\section{Data 140: Probability for Data Science}

Probability theory is the essential mathematical framework for formalizing the concepts introduced in Data 8 and its connector courses and providing a bridge to more advanced courses.  Data 140 is an introductory probability class for students who have taken Data 8, calculus, and linear algebra.

In most statistics departments, the traditional journey towards inference and data analysis begins with a calculus-based probability class, typically with no computational component; then a mathematical statistics class that typically includes some data analysis; and then classes in linear models, time series, and so on.  Data 140 is novel in that it uses inference as motivation for studying probability, and computing as a primary tool along with math. The course preserves the mathematical rigor of classical probability courses and then goes further with the theory, using the math and computation to enhance and inform each other just as they do in research.

Interleaving mathematics and computing is a natural progression for students who have taken Data 8 and want to go deeper into inference.  Indeed, for them it can be frustrating to take a traditional all-calculus probability class where the computation, if any, is done on hand calculators.  Some students find themselves at sea in abstract mathematics and need computation as an anchor.  No matter that the skills needed in mathematics and computation are largely the same---logic, abstraction, modularity, precision, and creativity.  To these students, one of the worlds feels natural and the other alien.  But they still want a way to go further into the probability that they have seen daily in Data 8: sampling variability, random permutations, the bootstrap, and always distributions, distributions, distributions.

Data 140 offers them the way.  The course is able to move fast: many of the concepts have already been motivated and discovered empirically in Data 8.  A side effect of the pace is that students with strong math preparation are also opting for Data 140 over traditional probability courses, because they know it will take them further.

The course content includes the material covered in standard undergraduate probability classes and also some inference, both frequentist and Bayesian, so that students can move on to a class in statistical learning and decisions without having to take a semester of mathematical statistics in between.  This is ambitious, and is achieved by careful coordination of all the material each week: lectures, practice, homework, and lab.

In some weeks we build directly on the ideas introduced in Data 8.  For example, total variation distance is introduced in Data 8 as a reasonable and straightforward way of quantifying the difference between two distributions on the same finite set of categories.  In Data 140 homework, students derive the interpretation of total variation as the biggest difference in probabilities assigned by the two distributions.  In lab in the same week, they compute the distance between binomial distributions and their Poisson approximations, and thus have a sense of how good the approximations are. As an extension of the method, Data 140 students have examined fixed points and consecutive pairs of random permutations, and top-to-random shuffles of a deck.

Shuffles reinforce the point that calculations can quickly get too large even for computers to handle.  This point is made on the very first lecture, in the context of the Birthday Attack, so that students start off with respect for the mathematics and pleasure in it. By the end of the term, when the lab has them construct bivariate normal random variables from independent standard normal components, the lecture derives the multiple regression estimate as a projection, and the homework has them come up with the multivariate normal distribution of the estimated coefficients, students are at ease in the world of math as well as computing.

Labs take the students beyond what is typically covered in a first course in probability. Markov chains and the Metropolis algorithm are covered in a fairly standard way, and then students use the algorithm to break a substitution code.  The beta and binomial families are explored in the context of Bayesian estimation, and then students construct a Chinese Restaurant clustering process and use beta and binomial facts to study its properties.  But mainly, the labs help develop and reinforce the math.  In the words of a student, ``[L]abs were shockingly helpful in developing certain theory without me even noticing.''

Like all our data science classes, Data 140 attempts to develop a way of thinking, and indeed the word ``think'' appears repeatedly in student comments.  ``Great mix of theory and developing a probabilistic way of thinking,'' said a student. ``This course can change the way you learn and think,'' said another. ``The pace is demanding and pushes you to improve faster than you think is possible, but in the middle of the semester you will change.  Things will start to click and it is indescribable how satisfying that is.''  And another said simply, ``It taught me how to think in a manner that enables me to succeed.''

Data 140 has not caused a drop in enrollments in the probability courses that the campus has long offered at this level in the Statistics department and elsewhere.  Instead, it engages about 650 more students annually in probability theory.  Data 8 and its connectors have motivated a diverse population to study probability: top Data 140 students in recent semesters have come from Sociology and Nutrition Science as well as from Applied Math, CS, Data Science, and Economics.

The interest is also spurred by our students' growing realization of the importance of probability theory and the need to understand it well.  ``I realized I couldn't keep faking my way through probability,'' said a student, explaining why they took this class after doing just fine in the CS department's demanding machine learning class.  ``Definitely a must-take class for anyone interested in Computer Science or Data Science,'' said another, while their classmate cast a wider net: ``[T]his is the one class I would recommend everyone taking before they graduate Berkeley.''

\section{Data 102: Data, Inference, and Decisions}

Data 102 is a senior-level course where the focus is decision-making.  While decision-making is present throughout our course sequence, it steps up to center stage in Data 102.  We think of our Data 102 students as likely future thought leaders in industry, science, academia, and government, and we want them to understand that data science is not merely about processing data, or solely about studying phenomena through the lens of data analysis, but often it is about making real-world decisions.  Decisions can have consequences.  Accordingly, in Data 102 we emphasize examples from domains such as public policy, medicine, and commerce, where algorithms are increasingly being used to make decisions, and where human happiness can hang in the balance.  We emphasize the social, economic, and ethical aspects of decision-making.

We begin the course in a very traditional fashion, teaching the rudiments of statistical decision theory (in a style that would be familiar to Wald, von Neumann, and Blackwell).  We start with basic two-alternative hypothesis testing, and take the opportunity to distinguish between Bayesian and frequentist versions of hypothesis testing, relying on Data 140 to give students the mathematical concepts so that these distinctions can be made crisp.  But, as is our wont, the focus is not the mathematics, but rather the thinking styles associated with the two perspectives.  Bayesian and frequentist perspectives are rarely brought together in undergraduate statistics curricula (not to mention graduate curricula), and when they are it is generally in the context of estimation, where Bayesian and frequentist approaches are more similar than different, and where the goal is to emphasize the similarities.  A sharper understanding is obtained by considering the contrast in thinking styles in the context of hypothesis testing.

We see the situation as being akin to that of quantum mechanics, where waves and particles provide related---but different---perspectives on physical phenomena.  A physics curriculum that simply picks one or the other, or merges the two in some vague way, would not be viewed as satisfactory.  Similarly, a data science curriculum needs to give some indication that the underpinnings of inference and decision-making are conceptually nontrivial, and are supported by two underlying, complementary perspectives.

To have this philosophical discussion yield something concrete, we turn the class in a 
rather non-traditional direction.  We explain that classical hypothesis testing focuses on 
a single decision, and we argue (via examples) that real-world decisions are rarely made 
in isolation.  Instead, decisions are made in the context of other decisions and in the 
context of other decision-makers.  We accordingly turn to \emph{multiple hypothesis 
testing}, and we introduce the error-rate criterion of \emph{false-discovery 
proportion} (FDP) as a way to measure performance when there are many hypotheses, 
distinguishing it from the zoo of classical criteria such as Type I and Type II 
errors, specificity, sensitivity, etc.---all of which are fine for single hypothesis 
tests, but are less useful for multiple hypothesis tests.  The FDP is a very natural 
quantity---it is simply the number of false discoveries divided by the total number 
of discoveries over a set of hypothesis tests.  It is natural enough that it's not 
hard to convince students of the practical value of such a criterion---they can well 
imagine a future boss asking them how many discoveries they've made today, and what 
fraction of those discoveries are worth investing in.  We also emphasize that the 
FDP is fundamentally a Bayesian quantity---it involves conditioning on the decisions, 
which depend on the data, instead of conditioning on the unknown truth.  We then ask 
a frequentist question---can we develop algorithms that keep the average FDP---the 
\emph{false discovery rate (FDR)}---below some desired value?

The cognoscenti will know that there exists such an algorithm, and that the algorithm 
has an appealing computational side to it, involving sorting and comparisons.  But we 
have a different didactic goal in mind at this point---the modern problem of \emph{online 
FDR control}.  The online FDR problem involves developing algorithms that maintain 
FDR control not only at the end of a batch of tests, but at any time along the way.  
Wearing our computational hats, we find that this perspective connects better to real 
software deployments of multiple testing, particularly in industry.  And, importantly,
a beautiful fact arises: there exist algorithms that are not only simple, but which
also have an elementary proof of online FDR control.  Indeed, it is a one-slide proof, 
perfectly suited to a senior-level undergraduate course.\footnote{For a recent treatment of online 
FDR algorithms and their theory, see \cite{ramdas2018saffron}.} 

We wish to emphasize two points about this discussion.  First, in developing our data science curriculum, we have often found that topics that are at the research frontier are often easier to motivate and explain than classical concepts, precisely because they have been developed in response to contemporary real-world problems, and because they make use of modern tools, most notably the computer.  Second, although we have de-emphasized mathematics in this article, we want to make clear that we are not against mathematics.  Quite to the contrary!  But we want the mathematics to support the concepts, rather than to be the concepts.  And we want the proofs that support the class to be, in the words of Paul Erd\"os, ``Proofs from the Book.''  Students should marvel at how the mathematics expresses powerful ideas simply and sharpens the mind.

In the design of Data 102, we continued to rely on ``vignettes'' to drive our thinking.  In particular, in the current version of the course, the COVID-19 pandemic is omnipresent in the lives of both students and teachers, providing a daily reminder of what consequential, on-line, societal-scale decision-making with uncertain data can look like. Accordingly, we take real-life examples from problems that we find in the newspaper, such as ascertaining the prevalence of infection, estimation of the case-fatality rate, and the design of clinical trials for vaccines.  Concepts such as the sensitivity and specificity of a diagnostic test take on real meaning for students in this context, and we are able to emphasize that these classical diagnostics---which are independent of the prevalence of the disease---yield decision rules that are very different from the FDR-based decision rules, which take prevalence into account.  Students can see that this really matters when prevalence is low, as it generally is (hopefully!) in an epidemiological context.  We can therefore contrast individual decision-making with group decision-making and discuss the practical and ethical implications of such a contrast for public policy.  We can also discuss the use of Markov-chain simulations of viral spread and how to make inferences based on such computations.  Overall, the COVID-19 examples allow us to drive home the point that one important mission for data science is to design and deploy societal-scale systems that help humans cope with emerging challenges.

Another vignette that helped to drive the course design involved combining an economic perspective on decision-making with a statistical perspective. In particular, we bring the economic concept of a \emph{matching market} together with the statistical concept of a \emph{multi-armed bandit}.  The teaching of matching markets allows us to introduce the famous Gale-Shapley algorithm, and to reason about its properties, as one would do in a classical computer science class.  To combine such algorithmic thinking with inferential thinking, we note that in real life, agents don't necessarily know their preferences a priori, but need to accumulate experience with the outcomes available in the market in order to \emph{learn} their preferences.  Multi-armed bandits provide an excellent example of exactly such a learning problem.  They are based on the use of confidence intervals (already taught in Data 140) to guide choices of which arm to pull in each of a sequence of trials. Putting these ideas together yields a concept that has only recently appeared in the research literature---a matching market in which each agent forms confidence intervals that are used by the Gale-Shapley algorithm to help a group of agents make decisions~\citep{Lydia}.  Students find this example not only mathematically interesting but also compelling in terms of its natural real-world applications.

The concept of \emph{private data analysis} provides fodder for additional vignettes.  Briefly, research in the theoretical computer science and database communities has given rise to the concept of \emph{differential privacy}, which provides algorithmic methods for adding noise to a database, yielding a ``privatized database'' that guarantees the privacy of individuals who supply data to the database while still allowing queries to be answered (nearly) correctly.  It is natural to supplement this computational story with an inferential story---how might we add noise to the database to not only ensure privacy but also to ensure inferential accuracy?  That is, will a privatized database make good predictions for individuals who come from the same population as those in the original data?

Finally, we also cover causal inference in Data 102.  In doing so, we close a loop from the senior experience back to the freshman experience---one of the first lectures in Data 8 is on causal inference, where we discuss John Snow's discovery of the cause of a cholera outbreak in London.  In Data 102, we discuss how to design experiments so that causal hypotheses can be evaluated, and we discuss inferring cause from observational data.  The latter is precisely what John Snow did in 1854, where, in so doing, he showed vividly how data analysis can be used to make consequential decisions and change the world for the better.

\section{Data 104: Human Contexts and Ethics of Data}

Recall that one of the major learning goals for the curriculum is that students should be ``systematically attentive to the social, cultural, and ethical contexts of the problems that they are formulating and aiming to solve.'' Accordingly, in all of our courses, we aim to tie our framing and our examples to plausible, real-world settings, and to provide explicit consideration of the interplay between social contexts and the increasingly wide deployment of data science methodology. Such contextualization is aided by the connector model, as well as by the domain emphasis that each data science major chooses to complete. But the lecture format of our technical courses falls short of achieving such a learning goal, and it was felt essential that the curriculum draw the threads together with a full course that gives students structured experience in grappling with the social and ethical ramifications of data collection, data analysis, and algorithmic decision-making, including their own work. Data 104, ``Human Contexts and Ethics of Data,'' provides such an experience. It is typically taken by students in their junior or senior year.

Data 104 aims to equip students to recognize, analyze, and take considered action in situations where data analysis, social context, and human welfare intersect. It does this by providing a conceptual framework anchored in disciplines that systematically examine human experience and society and by engaging students through reading, discussion, and writing. The structured framework helps students ``connect the dots'' between examples they see in the rest of the curriculum, where questions of representation, classification, prediction, decision-making, and justice are often at stake. In Data 104, broader concepts and methods from the social sciences and humanities are built into a scaffolding that students practice using across cases past and present. This sets them up for future contexts where circumstances will likely arise around data science technologies and systems that we have not even foreseen today.

Philosophically, Data 104 draws its understanding of ethics from the modern hermeneutic tradition, as captured in Paul Ricoeur's concept of ``aiming at the 'good life' with and for others in just institutions''~\citep{Ricoeur}. The course asks students to grapple with how human action toward this end is mutually constituted with the technologies that people design and use~\citep{Jasanoff}. These are abstract, open-ended questions, and our overall framing of human contexts and ethics comes from the research frontier, principally from philosophy, history, and STS (Science, Technology, and Society), including perspectives from sociology, anthropology, political theory, literature, and law. But in line with the rest of the curriculum, Data 104 brings its principles down to an undergraduate level and teaches them by working through concrete real-world cases. Students practice applying a Human Contexts and Ethics (HCE) toolkit, which brings together a set of core diagnostics to clarify the stakes of complicated situations and identify opportunities for action.

For instance, students get practice studying sociotechnical systems, hybridity, and agency in connection with automated decision-making for self-driving cars, using M.\ C.\ Elish's framing of a ``moral crumple zone'' to ask where responsibility gets assigned when systems break down~\citep{Elish}. Similarly, students work through positionality, representation, and power in relation to data governance, biomedical knowledge production, and structural inequity in the case of the Pima Indian Diabetes Dataset, which was derived from a longitudinal epidemiological study of Akimel O'odham people in the U.S. Southwest and is now part of a standard machine learning repository (Radin, 2017). The toolkit gives students a set of structured ways to contemplate different dimensions of a data science problem and suggest directions in which it is possible to take action. Because it focuses on opportunities for action in social context, it provides a complementary perspective to classical engineering ethics approaches that center on duties, consequences, or character. This framework lets students contextualize topics that appear in their other data science courses, such as data provenance, privacy, category construction, uncertainty, bias, and legacies of historical social inequalities in training datasets. The toolkit can be integrated into each phase of the data science lifecycle and can open up shared examples across the curriculum, such as the COMPAS algorithm and fairness, as well as whatever new cases students experience in their own work.

Data 104 also helps students understand how they and their experiences fit in a temporal and social landscape. The course gives them an introduction to the real-world historical dynamics that have spurred the growth of the data science profession and the transformational expectations placed on its technologies. The course's historical arc stretches from the real-world origins of statistics, to the emergence of questions about ethical practice in research and technology, to the drive for datafication in contemporary Silicon Valley. This framing aims to help students position themselves in the present and into the future with a fuller understanding of their world's underlying dynamics. The historical narrative also helps them put their cases into an ongoing flow of time. Alongside its consistent backbone of examples, Data 104 updates some fraction of its cases each semester, based on what is currently happening or what is on students' minds. In its most recent offering, it shone a spotlight on racial justice, exploring the long trajectory of surveillance of minoritized communities and the systematic inequalities of structural racism that have led to recurring racial biases in many algorithmic tools~\citep{browne,benjamin}.

Data 104 focuses students' effort on asking good questions, engaging with multiple perspectives, and making judgments that align with their own commitments. It aims to be student-centric in its reflection exercises, structured discussion, and written assignments. Students inquire into their own experiences, commitments, and visions of a good future. By design, weekly discussion sections mix the perspectives of data science majors and minors with students from the social sciences, humanities, pre-medical and pre-public health tracks, and other engineering disciplines. The course's final assignment is a ``vignette,'' in this case on a topic that each student chooses. The vignette invites them to reflect on a real-world situation, identify key aspects needing attention, and address an argument to their reader with the goal of shaping responsible action. Data 104 students' vignettes have dealt, for instance, with data science challenges to the framework of the Genetic Information Non-Discrimination Act, the possibilities for companies to design AI-enabled personal digital assistants with consent and transparency built in, and the tension between the need for more inferences about transgender populations and the unwarranted exposure this creates. As one student commented, ``I loved that this was a practical class---one that was rooted in PRACTICING the tools that we learned! The tangible vignette takeaway was also super cool, especially for data science people who may otherwise feel that they should leave this sort of thinking to social science folks.''

Just as in other parts of the curriculum, when we combine a systematic conceptual structure with praxis relevant to the real world, the combination draws students in. Students completing the course have remarked on their sense of clarity, self-recognition, and career empowerment. ``I plan to go into the tech industry one day, so this social side of issues affecting not only data but communities of color and people of different backgrounds like myself is important for me to learn from and understand,'' noted an applied mathematics/data science double major. ``The course helped me choose my career path out of undergrad,'' commented another data science student, ``to join a smaller, tighter knit team where I would have more opportunity to set good practice. I think I was aware of a lot of issues, but could never really put it in words, and now I have the vocabulary to speak about it.'' As a data science/computer science double major reflected, ``it makes me take my decisions in data science much more seriously and mindfully, also gives me the courage to speak up.''

We believe that the new synthesis of computation and inference that runs throughout Berkeley's data science curriculum requires new thinking about human contexts and ethics. Classical problems of data ethics (for instance, consent, privacy, and security) and responsible statistical practice need to be complemented and extended to account for the extraordinary reach of computing---how it is deployed to automate processes, digitize and quantify the world, connect data and people, and shape human behavior on computational platforms. The fusion of computing and inference in automated decision-making systems raises the stakes and puts existing challenges in a new light. This makes it all the more essential that our students be prepared to engage with these dynamics not just as we see them today, but as they will emerge in the future. Underneath all of our efforts at providing a conceptual framework are crucial questions about who frames the problems whose solutions are sought and the larger dynamics of our social world.

\section{Uptake and Engagement}

Almost 3,000 UC Berkeley students took Data 8 in the 2019-2020 academic year.  We expect that nearly 50\% of the most recent class of UC Berkeley undergraduates will enroll in Data 8 before they graduate. Its popularity among students prompted departments across campus to consider Data 8 as a potential part of their degree programs, and it now fulfills the statistics requirement for 25 different major programs, from Civil Engineering to Legal Studies to Public Health. In a time when demand for computing and statistics education is surging, Data 8 plays a critical role in providing broadly accessible and relevant instruction to the UC Berkeley undergraduate population.  Enrollment in undergraduate computer science lecture courses at UC Berkeley has increased by 453\% in the last ten years.\footnote{From 3,633 enrollments in the 2008-2009 academic year to 20,079 enrollments in the 2018-2019 academic year.}  Meeting this new demand requires more than just scaling up existing courses; we believe that Berkeley and other institutions must create educational pathways that are relevant to the very broad set of academic interests represented by the enormous cohort of students who have identified computational thinking and inferential thinking as core skills for the 21st century.

The diversity of participation in Data 8 is noteworthy---last year, 52\% of students who completed the course were female and 11\% were from underrepresented groups. To foster and retain their interest, and help develop them as leaders, the Data Scholars program offers support for the students' entire data science trajectory starting with Data 8. Modeled on Berkeley's highly successful Biology Scholars and CS Scholars programs, the Data Scholars program is designed for students from underrepresented minority backgrounds and serves about 80 students per semester. It includes academic support for students in Data 8 and for those who continue on to more advanced classes in the Data Science major, additional mentoring and guidance for those who participate in research, pedagogical guidance for those aiming to join the staff in large classes, and career support for those who want to use data science in their professions. Increasing numbers of students on the Data Scholars support teams have themselves been Data Scholars.

Data 8 has extended beyond UC Berkeley through offerings at other institutions and through Data 8X, a massive open online course sequence on the EdX platform. Faculty from over 200 peer educational institutions have reached out to the Data 8 team about offering the course at their colleges and universities. Several external versions of the course have been taught already, and many more are under development. The initial offering of Data 8X attracted 50,000 learners globally, and another 40,000 learners have enrolled in subsequent offerings. All course materials are freely available for both students and instructors, including the online textbook, the assignments and projects, and guides for setting up the course infrastructure so that students can complete their assignments, and instructors can assess them efficiently at scale.

Students routinely use the content of data science classes to answer questions that arise outside class. This tradition arose when permutation tests first appeared in the pilot offering of Data 8---by the next day, students had applied the test to letter grade data for a CS course taught by multiple faculty and had declared one of the professors to be ``the hardest.'' In response to numerous reports of students using the methods learned in data science classes in their own clubs, in other classes, and in internships, Berkeley's Division of Computing, Data Science, and Society (CDSS) created a formal structure for undergraduate research. The Discovery Research Program includes about 200 undergraduates every semester, in research projects guided by faculty across campus, or in projects of their own, supervised by a CDSS faculty member. The projects vary widely in the amount of technical expertise required: many require only Data 8. About 10 to 20 students each semester support the faculty teaching Data 8 connector courses. About a dozen students work on the Modules team with campus faculty who wish to introduce data science examples in their courses.

Data 100 is the largest upper-division course on the Berkeley campus with a steadily growing enrollment that topped 2,500 students last year. The course serves as a core upper-division requirement for two new degree programs at Berkeley: the data science major introduced in 2018 and the data science minor introduced in 2019. The course is also popular among students majoring in computer science, statistics, and engineering. Variants of Data 100 have been offered at other campuses in the University of California system, and we expect that the many institutions adopting Data 8 will eventually adopt Data 100 as well.

Data 104, Data 140 and Data 102 were designed specifically for students majoring in data science. Even these specialized courses are attracting hundreds of students each semester. For students to specialize in making data-driven decisions as their vocation, they must have an opportunity at the undergraduate level to develop a deep technical understanding of uncertainty, inference, and the implications of carrying out inference on data represented by a computer. They must simultaneously develop a sophisticated perspective on the social and ethical implications of these practices.  The curricula of these three courses are advanced and ambitious precisely in support of this goal---to develop a generation of thought leaders in data science.

\section{Teaching at Scale}

A new model of large-scale course delivery was required to offer Data 8 to a large fraction of Berkeley's undergraduate population. Not only is the absolute number of students enrolled in the course very large, but there is tremendous variety in the academic interests and intended majors of these students. When subjects reach this level of popularity, they often branch into population-specific variants. For example, Berkeley has three different calculus sequences: one for biology, one for physical sciences and engineering, and a third for everyone else. The variety of introductory statistics courses is greater still, including discipline-specific courses in public health, linguistics, and more. By contrast, Data 8 was designed to provide a single common core that is both accessible and challenging to a broad cross-section of undergraduates on campus. Students learn about connections to their specific domains of interest through the combination of connector courses and a team of teaching assistants from departments across campus. The structure of the teaching staff and the design of assignments are both intended to allow students to conduct ambitious data analysis projects within their first semester, even those without prior experience in statistics or programming. The Data 8 model is meant to support students who will go on to specialize in data science as well as those who will choose to focus on a data-rich domain within the natural sciences, social sciences, engineering, or data-focused humanities.

Many of the connector courses offered today were developed and offered during the same year that Data 8 was originally piloted. Indeed, Data 8 and its connector courses have been viewed from the outset as components in a unified learning experience for students that is both essential in its broad applicability and specific in its alignment with the particular academic interests of students. Developing this diverse suite of courses in a way that maintained their coherence required a substantial internal recruiting effort to identify faculty who would develop connector courses, as well as an unusually high degree of coordination among course instructors to ensure that the connector courses did in fact connect with the material in Data 8. At Berkeley, this coordination was enabled in large part because the intention to develop Data 8 and its connector courses was not isolated to a particular department or small group of faculty, but instead was cultivated by a diverse faculty committee convened by the chancellor to create a campus-wide unified approach to data science education. Before any particular course syllabus was ever created, stakeholders from multiple departments, divisions, and schools were invested in creating this new course experience. The team of instructors who created both Data 8 and its connector courses came together under the guidance of this committee who first identified that a common intellectual core existed among so many different disciplines.  Individual courses can be created by individual faculty members, but creating a coherent education program requires the commitment and energy of many.

One organizational component that proved critical was that someone who was not responsible for the day-to-day development and delivery of Data 8 took responsibility for ensuring that the Data 8 and connector instructors met regularly, shared their plans, and kept each other apprised of what students were learning in each of the courses. This coordination effort led to the development of a summer seminar for faculty that was so popular among Berkeley faculty that it has now expanded to become the National Workshop on Data Science Education that convenes faculty from around the world to discuss course content and data science pedagogy.\footnote{\url{https://data.berkeley.edu/education/data-science-education-opportunities}} 

Another fruitful coordination tactic has been to involve undergraduates who have taken Data 8 as tutors and teaching assistants for connector courses. Because they have recently taken Data 8, they often know the material and course cadence well enough to help connector course instructors identify connection points between their course and Data 8.  

Undergraduates are also heavily involved in delivering the weekly lab sections of Data 8 at Berkeley, as well as helping students complete assignments through various forms of tutoring. Large-scale delivery of an introductory course, especially a new course such as Data 8, can benefit greatly from a robust undergraduate teaching program in which students first gain experience and confidence in support roles, then take on broader, more autonomous, and more essential roles as they mature. Teaching mentorship, either within a course staff or through a pedagogy course, can help with recruiting and developing more capable undergraduate tutors and teaching assistants. Although more logistically complicated, there are appealing advantages to building a course staff with a large number of undergraduates each working for a small number of hours each week. Undergraduate teaching roles with a lower time commitment are less likely to interfere with academic work, and a higher ratio of teachers to students can promote more individualized instruction. While the needs of each course are different, undergraduate tutors and teaching assistants are also heavily involved in teaching Data 100 and Data 140 alongside graduate students (but Data 102 and 104 are staffed primarily by graduate students).

To give a sense of scale---in its most recent incarnation in 2020, Data 8 had close to 90 students on the paid instructional staff as teaching assistants and tutors.  Offering a data science curriculum at scale requires student involvement at scale.

For all of our courses, the student experience outside of lecture is a central concern. A common feature across Data 8, Data 100, and Data 140 is a weekly section in which students work through example problems using the same computing environment in which they will complete assignments. While these lab sections typically do not introduce new concepts or techniques, they do introduce new applications, new ways of combining ideas from the course, and new problem-solving strategies. This design is meant to ensure that students spend their time productively throughout the course: lab time offers review and reinforcement but minimizes redundancy with lecture, and the experience of solving problems in lab prepares students to solve similar problems on assignments and exams. Many students still need assistance in completing assignments, and so various forms of free one-on-one and small-group tutoring are available to students in all of these courses so that they can continue to make progress on assignments in a timely manner when they are unable to make progress on their own. The need for technical assistance during the course is minimized by using hosted Jupyter notebooks that are configured and maintained centrally. As a result, students can spend more time on course material rather than software configuration, and students who have less computing experience are not put at a disadvantage because of their lack of familiarity with software development tools.

There are also issues with computational infrastructure that have arisen as our enrollments have grown.  Indeed, offering large-scale courses that provide students with a hosted computing environment requires an investment in both software administration effort and cloud computing resources. The team that manages this effort at Berkeley has published a guide that describes the components of the course software stack and how to administer them~\citep{Nakagawa}.  UC Berkeley has been fortunate to receive resource support from industrial cloud computing providers to offset the cost of supporting students' computational requirements across many of our courses.

Finally, we note that numerous other colleges and universities have adopted some parts of the Berkeley curriculum and every new setting has revealed new educational challenges. Most notably, while we believe that the material in Data 8 can and should be within the reach of any college freshman, an aspiration that may be realized fully only after high schools have had time to react to emerging new curricula, we also acknowledge that the Berkeley students may have had more preparation than other populations. For different student populations, it could make sense to spread the content of Data 8 across two academic quarters or semesters, perhaps supplementing the material with extra practice beforehand and along the way.

\section{Coda}

The Berkeley Data Science curriculum arose as a pedagogical leap of faith---a belief that computation and inference are natural allies and they should be taught together, as a blend, in a modern undergraduate curriculum.  Having watched developments at the research frontier in both computer science and statistics over many years, we had been struck by the fact that the former has become increasingly probabilistic and inferential, and the latter has become increasingly computational, and that these developments have led, perhaps surprisingly, to ideas that have often been simpler to understand and to deploy than their classical counterparts.  But there really was no surprise---the classical insistence in computer science on determinism and the classical insistence in statistics on analytical results imposed severe limits on both fields, constraining the scope of their applications and requiring cleverness on the part of practitioners to surmount those limitations in real-world applications.

Pedagogy in both fields often aimed at bringing young minds to heel within the classical frameworks, while only occasionally admitting the limitations of those frameworks.  Powerful ideas that blend probability, inference, and computation were reserved for PhD-level seminars and research 
conferences---undergraduates were kept in the dark.

While the emergence of data science on the academic landscape at many universities has led to discussions of curricular change, these discussions have often focused on the mere juxtaposition of existing curricula in computer science and statistics.  Such a juxtaposition comprises a great deal of material, and the ensuing discussion has often focused on what to omit.  A problem with such a framing, in our view, is that the classical curricula were overly disjoint and the opportunities for blending were difficult to perceive.

We believe that data science provides a historical opportunity to concoct a new blend, transporting powerful ideas from the research frontier into the undergraduate curriculum, and empowering students to pose and solve a broad range of emerging problems.  We believe that we can and should expose a new generation of students to intellectual foundations that are simultaneously computational and inferential.  We believe that students will perceive the simplicity and generality of this combination and will view it as a foundation on which to build their own thinking and solve their own problems.

Taking our enthusiasm to the limit, we would like to end with the following provocative question: Should the new data science curriculum entirely displace classical curricula in computer science and statistics, or should it simply live side by side with those curricula?

Of course, the new curriculum comprises many classes in addition to those that we have discussed here, and those classes are drawn in part from the classical curricula, so there is necessarily an overlap. The question that we are asking is whether the main architectural elements of the curriculum---the stage-setting introductory class and the main backbone of the curriculum---are those that we have discussed here or their counterparts in a classical computer science or statistics curriculum. We offer a few opinions on this question, with the goal of opening a debate, not closing it.

Certainly there are elegant, deep, and powerful ideas in classical curricula that should continue to be nurtured.  We have in mind notions of \emph{abstraction} in computer science, \emph{sampling} in statistics, and different flavors of \emph{modeling} in both disciplines.  The new curriculum certainly teaches these notions, but the range of problem domains in which they are developed is arguably narrowed.  For example, a data-science student will likely have less exposure to logic-based abstractions in computer science that underpin the fields of verification and cryptography.  A data-science student may have less exposure than a classical statistics student to the range of applications of analysis of variance.

But overall, we would have confidence in the ability of one of our data-science students to compete with classically-trained peers in either industry or academia. Our data-science students will be able to write code, but they will do so with an enhanced perception of the real-world consequences of that code; in particular they will be aware of the fact that code often operates in a stochastic environment, and conclusions are always tentative.  Moreover, their education will have in it the seeds of a broad range of mathematical ideas that can be further developed in a PhD.

One useful perspective on this debate comes from considering the idea of a ``fifth year.'' Education is not merely about exposure to a certain set of ideas; it is about the intellectual maturity that comes about from the lived experience of working with those ideas, in the context of a range of problems, and in the context of one's own life experience.  Mature understanding of deep concepts such as abstraction, sampling, and modeling takes time.  One can imagine a fifth-year program that is project-based, providing time for maturity to emerge, introducing a wider range of concepts from classical curricula, and continuing to emphasize engagement with fields beyond computer science and statistics.

\subsection*{Acknowledgments}
We would like to thank the many colleagues at Berkeley who have been our partners over the past few years as the Berkeley Data Science curriculum has been rolled out, from faculty who helped to design and teach the classes, to graduate students who have played key roles as teaching assistants and mentors, to staff who have helped to build the infrastructure behind the curriculum, to the departments who came together as stakeholders in a campus-wide initiative, but most of all to the thousands of undergraduate students whose energy, passion, open minds, creativity, and courage has inspired us and challenged us.  Their excitement at doing something new and impactful, and their willingness to help each other and to help us, has confirmed our abiding faith in liberal education.

\bibliographystyle{apalike}
\bibliography{adj}

\begin{thebibliography}{}

\bibitem[Abelson and Sussman, 1985]{sicp}
Abelson, H. and Sussman, G.~J. (1985).
\newblock {\em Structure and Interpretation of Computer Programs}.
\newblock The MIT Press, Cambridge, MA.

\bibitem[{ACLU of Northern California}, 2010]{aclu2010}
{ACLU of Northern California} (2010).
\newblock {\em Racial and Ethnic Disparities in Alameda County Jury Pools}.

\bibitem[Benjamin, 2019]{benjamin}
Benjamin, R. (2019).
\newblock {\em Race after Technology: Abolitionist Tools for the New Jim Code}.
\newblock Polity, Medford, MA.

\bibitem[Browne, 2015]{browne}
Browne, S. (2015).
\newblock {\em Dark Matters: On the Surveillance of Blackness}.
\newblock Duke University Press, Durham, NC.

\bibitem[Cobb, 2015]{cobb2015}
Cobb, G. (2015).
\newblock Mere renovation is too little too late: We need to rethink our
  undergraduate curriculum from the ground up.
\newblock {\em The American Statistician}, 69(4).

\bibitem[De~Veaux et~al., 2017]{deveaux}
De~Veaux, R.~D., Agarwal, M., Averett, M., Baumer, B., Bray, A.;~Bressoud,
  T.~C., Bryant, L., Cheng, L., Francis, A., Gould, R., Kim, A., Kretchmar, M.,
  Lu, Q., Moskol, A., Nolan, D., Pelayo, R., Raleigh, S., Sethi, R.~J.,
  Sondjaja, M., Tiruviluamala, N., Uhlig, P., Washington, T., Wesley, C.,
  White, D., and Ye, P. (2017).
\newblock Curriculum guidelines for undergraduate programs in data science.
\newblock {\em Annual Review of Statistics and Its Application}, 4:15--30.

\bibitem[Elish, 2019]{Elish}
Elish, M.~C. (2019).
\newblock Moral crumple zones: Cautionary tales in human-robot interaction.
\newblock {\em Engaging Science, Technology, and Society}, 5:40--60.

\bibitem[Fisher, 1935]{Fisher}
Fisher, R.~A. (1935).
\newblock {\em The Design of Experiments}.
\newblock Hafner Publishing Co., New York.

\bibitem[Franklin, 1999]{Franklin}
Franklin, U. (1999).
\newblock {\em The Real World of Technology}.
\newblock House of Anansi Press, Toronto, Canada.

\bibitem[Freedman et~al., 1978]{fpp}
Freedman, D., Pisani, R., and Purves, R. (1978).
\newblock {\em Statistics}.
\newblock W. W. Norton, New York.

\bibitem[Jasanoff, 2004]{Jasanoff}
Jasanoff, S. (2004).
\newblock {\em States of Knowledge: The Co-production of Science and Social
  Order}.
\newblock Routledge, New York.

\bibitem[Liu et~al., 2020]{Lydia}
Liu, L., Mania, H., and Jordan, M.~I. (2020).
\newblock Competing bandits in matching markets.
\newblock In {\em Proceedings of the Twenty-Third Conference on Artificial
  Intelligence and Statistics (AISTATS)}, Palermo, Italy.

\bibitem[Nakagawa et~al., 2020]{Nakagawa}
Nakagawa, A., Liang, A., Lim, R., Usovich, K., Pyles, C., Mehendale, A., and
  Holdgraf, C. (2020).
\newblock {The Data Science Educator's Guide to Technology Infrastructure}.

\bibitem[Pitman, 1937]{Pitman}
Pitman, E. J.~G. (1937).
\newblock Significance tests which may be applied to samples from any
  populations.
\newblock {\em Royal Statistical Society Supplement}, 4:119--130.

\bibitem[Ramdas et~al., 2018]{ramdas2018saffron}
Ramdas, A., Zrnic, T., Wainwright, W., and Jordan, M.~I. (2018).
\newblock {SAFFRON}: an adaptive algorithm for online control of the false
  discovery rate.
\newblock In {\em Proceedings of the 35th International Conference on Machine
  Learning}, Stockholm, Sweden.

\bibitem[Ricoeur, 1992]{Ricoeur}
Ricoeur, P. (1992).
\newblock {\em Oneself as Another, Trans. Kathleen Blamey}.
\newblock University of Chicago Press, Chicago, IL.

\end{thebibliography}

\end{document}